\documentclass[english]{cccconf}
\usepackage[comma,numbers,square,sort&compress]{natbib}
\usepackage{epstopdf}

\usepackage{longtable}
\usepackage{subfigure}
\usepackage{amsmath}
\allowdisplaybreaks
\usepackage{color}
\usepackage{footnote}
\usepackage{algorithm}
\usepackage{algpseudocode} 
\usepackage{algpseudocode}
\usepackage{algorithmicx}
\usepackage{multirow} 
\usepackage{booktabs}
\usepackage{graphicx}
\usepackage{amssymb}
\usepackage{amsbsy}
\usepackage{array}
\usepackage{longtable}
\usepackage{epstopdf}
\usepackage{pbox}
\usepackage{breqn}
\usepackage{mathrsfs}
\usepackage{multicol}
\usepackage{supertabular}
\usepackage{enumerate}
\usepackage{url}
\usepackage[justification=centering]{caption}

\newcommand{\sgn}{\mathrm{sgn}}

\begin{document}
\title{Learning-Based Adaptive Dynamic Routing with Stability Guarantee for a Single-Origin-Single-Destination Network}

\author{Yidan Wu\aref{JI},
        Feixiang Shu\aref{SEIEE},
        Jianan Zhang\aref{PKU},
        Li Jin\aref{UMJI}}
\affiliation[JI]{UM Joint Institute, Shanghai Jiao Tong University, Shanghai, China
        \email{wyd510@sjtu.edu.cn}}
\affiliation[SEIEE]{School of Electronic Information and Electrical Engineering, Shanghai Jiao Tong University, Shanghai, China
        \email{sfx-sjtu@sjtu.edu.cn}}
\affiliation[PKU]{School of Electronics, Peking University, Beijing, China
        \email{zhangjianan@pku.edu.cn}}
\affiliation[UMJI]{UM Joint Institute and School of Electronic Information and Electrical Engineering, Shanghai Jiao Tong University, Shanghai, China
        \email{li.jin@sjtu.edu.cn}}
\maketitle

\begin{abstract}
We consider learning-based adaptive dynamic routing for a single-origin-single-destination queuing network with stability guarantees. Specifically, we study a class of generalized shortest path policies that can be parameterized by only two constants via a piecewise-linear function. Using the Foster-Lyapunov stability theory, we develop a criterion on the parameters to ensure mean boundedness of the traffic state. Then, we develop a policy iteration algorithm that learns the parameters from realized sample paths. Importantly, the piecewise-linear function is both integrated into the Lyapunov function for stability analysis and used as a proxy of the value function for policy iteration; hence, stability is inherently ensured for the learned policy. Finally, we demonstrate via a numerical example that the proposed algorithm learns a near-optimal routing policy with an acceptable optimality gap but significantly higher computational efficiency compared with a standard neural network-based algorithm.
\end{abstract}


\footnotetext{This work was in part supported by National Natural Science Foundation of China Project 62103260, SJTU UM Joint Institute, J. Wu \& J. Sun Foundation.}

\section{Introduction}
\label{sec:intro}

Dynamic routing is a control scheme with extensive application in network systems including communications, manufacturing, transportation, etc \cite{mitra1991comparative,down1997piecewise,alanyali1997analysis,jin2018stability}. A primary objective of dynamic routing is to ensure stability, i.e. the boundedness of the long-time average of traffic state. The typical tool to approach this class of problems is the Lyapunov function method \cite{kumar1995stability}. On top of this, a secondary objective is to minimize the system time experienced by each job. Queuing theorists have made efforts on characterization of the limiting distribution of the traffic state, but such results are usually complicated and non-scalable \cite{foschini1978basic,ephremides1980simple,tassiulas1992stability}. Recently, learning-based approaches have been proposed for dynamic routing. However, such approaches are usually purely numerical without explicit theoretical guarantees on traffic stability \cite{fadlullah2017state}.

\begin{figure}[!htb]
    \centering
    \includegraphics[width=\hsize]{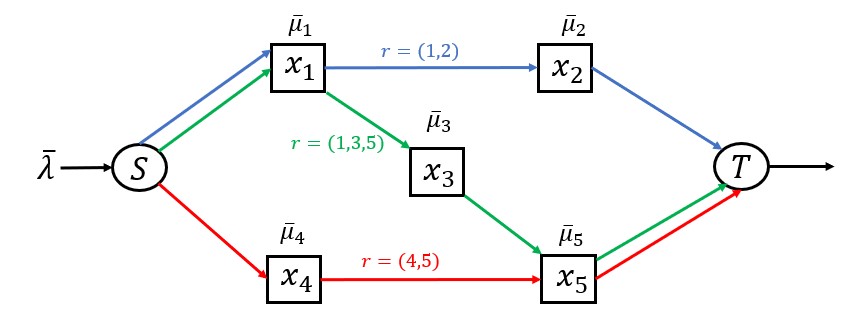}
    \caption{A single-origin-single-destination network.}
    \label{fig:network1}
\end{figure}

Motivated by the above challenge, we consider a dynamic routing scheme for single-origin-single-destination queuing networks that integrates Lyapunov function methods \cite{meyn_tweedie_1993} and approximate dynamic programming \cite{bertsekas2012dynamic}. 
We use a class of piecewise-linear (PL) functions that are increasing in traffic states to measure the travel cost of a path. An incoming job is routed to the path with the smallest cost, so the routing scheme can be viewed as a generalized shortest-path (GSP) strategy. We use the PL functions to construct a Lyapunov function for stability analysis, and we show that the resultant routing scheme is throughput-optimal. We also use the PL functions as proxies for action-value functions to obtain a policy iteration algorithm. This algorithm computes a suboptimal routing scheme that is guaranteed to be stabilizing. We also show that our scheme gives comparable performance (i.e., average system time) with respect to neural network-based methods but with a significant reduction in computational workload.
This paper focuses on the bridge network in Fig.~\ref{fig:network1},
but our modeling, analysis, and computation approaches also provide hints for general networks.

Extensive results have been developed for traffic stabilization and throughput maximization \cite{kumar1995stability,dai1995stability,hollot2002analysis,georgiadis2006resource}, which provide a basis for this paper.
In terms of travel cost minimization, previous results are typically either restricted to simple scenarios (e.g., parallel queues) or based on numerical simulation \cite{sinha2017optimal,zhang2021optimal} or optimization methods without traffic stability analysis \cite{ross1986optimal,li2019overview}. In particular, reinforcement learning methods are proposed to compute optimal/suboptimal solutions to dynamic routing problems \cite{xu2018experience,bernardez2021machine}. Although learning methods usually give excellent performance, they focus on algorithm design and do not emphasize theoretical guarantees for stability/optimality. In addition, generic learning methods may be computationally inefficient for dynamic routing. In particular, Liu et al. proposed a mixed scheme that uses learning over a bounded set of states and uses a known stabilizing policy for the other states \cite{liu2022rl}; this notion of integration of learning and stability guarantee is aligned with the idea of this paper, but this paper makes such integration in a different manner.

We formulate our problem on the single-origin-single-destination queuing network in Fig.~\ref{fig:network1}. Jobs arrive at the source node as a Poisson process. Each server has exponentially distributed service times. When a job arrives, the system operator assigns a path to the job, which will not be changed afterwards. Hence, every job belongs to a class that corresponds to the assigned path. The key of our routing scheme is a set of PL functions that approximate the expected travel times on each path. Such PL functions were introduced by Down and Meyn \cite{meyn_tweedie_1993} to show positive Harris recurrence of the traffic state. Our previous work proposed an explicit construction of the parameters of the PL functions according to network topology, without solving model data-dependent inequalities \cite{7}. In this paper, we further utilize the PL functions to bound the mean of the traffic state and to obtain our GSP routing policy.

We show that our GSP policy maximizes network throughput in the sense that it is stabilizing as long as the demand at the source is less than the min-cut capacity of the network (Theorem~\ref{theorem1}). The proof utilizes a critical insight about the PL functions: each linear regime corresponds to a particular configuration of bottlenecks. We construct a piecewise-quadratic Lyapunov function using the PL functions and show that the mean drift of the Lyapunov function is negative in heavy-traffic regimes. The drift condition also leads to a set of constraints on the PL parameters.

We then develop a policy iteration algorithm that learns the PL parameters (and thus the GSP policy). The policy iteration algorithm uses the PL functions as a proxy (as opposed to an immediate approximation) of the action-value functions; this connects the learning-based routing decisions with the above-mentioned theoretical properties. In each iteration, the current policy is evaluated using the PL function after a Monte-Carlo episode and improved by greedifying over the PL functions; the evaluation is done by an algorithm specifically designed for the PL functions. We compare our policy iteration algorithm with a standard neural-network dynamic programming algorithm. The implementation results indicate that our approach can attain a 1.9\%-6.7\% optimality gap with respect to the neural-network method with a 97.8\%-99.8\% reduction in training time.


In summary, our contributions are:
\begin{enumerate}
    \item A simple, insightful, and easy-to-implement dynamic routing scheme, i.e., the GSP policy.
    \item Theoretical guarantee on traffic stabilization and throughput maximization of the GSP policy over the bridge network.
    \item An efficient policy iteration algorithm for the GSP policy that attains comparable control performance but significantly reduces training time compared with neural network-based methods.
\end{enumerate}

The rest of this paper is organized as follows. Section~\ref{sec_model} introduces the network model and formulates the dynamic routing problem. Section~\ref{sec_stability} develops the stability guarantee for the GSP policy. Section~\ref{sec_learning} demonstrates the training algorithm for the GSP policy. Section~\ref{sec_conclude} gives the concluding remarks.
\section{Modeling and formulation}
\label{sec_model}
\newtheorem{theorem}{Theorem}
\newtheorem{assumption}{Assumption}
\newtheorem{definition}{Definition}

Consider the single-origin-single-destination (SOSD) network of queuing servers with infinite buffer sizes in Fig.~\ref{fig:network1}.
Let $\mathcal{N}=\{1,2,3,4,5\}$ be the set of servers. Each server $n$ has an exponential service rate $\bar\mu_{n}$ and job number $x_n(t)$ at time $t$, $t \in \mathbb{R}_{\ge 0}$. Jobs arrive at origin $S$ according to a Poisson process of rate $\bar\lambda > 0$. There are three $paths$ from the origin to the destination, $\mathcal{P}=\{(1,2),~(1,3,5),~(4,5)\}$. We denote $n\in p$ if server $n$ is on path $p.$

When a job arrives, it will go to one of the three paths according to a routing policy. The path of the job cannot be changed after it is decided at the origin. Thus, we divide the jobs into three classes according to their assigned paths. We use $x_n^p(t)$ to denote the jobs queuing in server $n$ and going on path $p \in \mathcal{P}$ at time $t$. For each server we have $x_n(t)=\sum_{n\in p}x_n^p(t)$. Thus we can define the state of the network as the vector $x=[x_1^{12},x_1^{135},x_2^{12},x_3^{135},x_4^{45},x_5^{135},x_5^{45}]^T$, and the state space is $\mathcal{X}=\mathbb{Z}_{\ge 0}^{7}$. 

We consider two sets of actions, viz. $1)~routing$ and $2)~scheduling$. Routing refers to determining the path of each job upon its arrival, which only occurs at $S$. The routing action can be formulated as a vector 
$$
\lambda=[\lambda_p]_{p\in\mathcal P}=\left[\begin{matrix}
\lambda_{12}\\
\lambda_{135}\\
\lambda_{45}
\end{matrix}\right]
\in\{0,\bar\lambda\}^3.
$$
Scheduling refers to selecting a job from the queue to serve, which only occurs at Servers 1 and 5; preemptive priority is assumed at both servers. 
The scheduling action can also be formulated as a vector 
$$
\mu=[\mu_n^p]_{p\ni n,n\in\mathcal N}=\left[\begin{matrix}
\mu^{12}_1\\
\mu^{135}_1\\
\mu^{12}_2\\
\mu_3^{135}\\
\mu_4^{45}\\
\mu_5^{135}\\
\mu^{45}_5
\end{matrix}\right]
\in\prod_{n\in\mathcal N}\{0,\bar\mu_n\}^{m_n},
$$
where $m_n$ is the number of paths through server $n$.

This paper focuses on a particular class of control policies which we call the $generalized~shortest~path$ (GSP) policy, which is based on the following piecewise-linear (PL) functions of traffic states:
\begin{align*}
    & Q_{12}(x)=\text{max}\{\beta^2 x_1^{12}, \beta(x_1^{12}+x_2^{12})\},\nonumber\\
    & Q_{135}(x)=\text{max}\{\beta^2 x_1^{135},\beta(x_1^{135}+x_3^{135}), \\
    &\hspace{2.7cm}(x_1^{135}+x_3^{135}+x_5^{135})\},\nonumber\\
    & Q_{45}(x)=\text{max}\{\beta^2 x_4^{45}, \beta(x_4^{45}+x_5^{45})\},
\end{align*}
where $\beta>1$ is a design parameter. 
We then define the notion of bottlenecks as follows: 
\begin{definition}[Bottleneck]
\label{def:dominance}
    Given $x\in \mathcal{X}$, a server $n\in p$ is a \emph{bottleneck} if
    \begin{enumerate}
        \item $
\frac{\partial_+}{\partial x_n}Q_p(x)>0,
    $ where $\partial_+$ means the right derivative, and

    \item either $n$ has no downstream servers or the immediately downstream server $n'$ is such that $
\frac{\partial_+}{\partial x_{n'}}Q_p(x)=0.
    $
    \end{enumerate}
\end{definition}
Intuitively, a bottleneck is a server that immediately contributes to $Q_p$, while the immediately downstream server, if it exists, does not contribute to $Q_p$.
Let $B=(b_{12},b_{135},b_{45})$ be a combination of bottlenecks, where $b_{12} \in \{1,2\},b_{135}\in\{1,3,5\},b_{45}\in\{4,5\}$, and denote $n\in B$ if $n$ is a bottleneck in combination $B$.
Let $\mathcal{B}$ be the set of possible combinations, which has 12 elements, each corresponding to a subset of $\mathcal X$ over which $\{Q_p\}$ are all linear in $x$. 

The GSP policy essentially sends an incoming job to the path with the smallest $Q$ function. However, the $Q$ functions only capture the traffic states but do not consider the influence of the service rates, especially the service rate of the current bottleneck. 
To account for this, we introduce a second parameter $\gamma>1$ that allows the GSP policy to prioritize the path with a higher bottleneck service rate.
To be specific, for a combination $B\in\mathcal B$, we first sort the paths according to their bottleneck service rates:
\begin{align*}
    &P_{1,B} :=\{p\in\mathcal P:p\in\mathop{\arg\max}\limits_{n\in B\cap p'}\bar\mu_{n}\},
    \nonumber\\
    &P_{\gamma,B}:=\{p\in\mathcal P:p\in\mathop{\arg\max}\limits_{p\in B\cap p'\cap P_{1,B}^c}\bar\mu_{b_{p'}}\},\nonumber\\
    &P_{\gamma^2,B}:=\mathcal P\backslash(P_{1,B}\cup P_{\gamma,B}).
\end{align*}
Then for $B\in\mathcal B$, define the \emph{weight} of each path as
\begin{equation} 
\label{path gamma}
\gamma_p(B):=
\begin{cases}
1 &p\in P_{1,B},\\
\gamma &p\in P_{\gamma,B}, \\
\gamma^2 &p\in P_{\gamma^2,B},
\end{cases}
\quad p\in\mathcal P.
\end{equation}

Now we are ready to formulate the GSP policy, with a slight abuse of notation, as follows: When there are no ties,
\begin{align*}
    &\lambda_p(x)=\bar\lambda\mathbb{I}\Big\{p=\mathop{\arg \min}\limits_{p'\in\mathcal{P}} \gamma_{p'} Q_{p'}\Big\},\ p\in\mathcal P,\nonumber\\
    &\mu^p_n(x)=\bar\mu_n\mathbb{I}\Big\{p=\mathop{\arg \max}\limits_{p':n\in B\cap p'} Q_{p'}\Big\},\ n\in\{1,5\},\nonumber\\
    &\mu^p_n(x)=\bar\mu_n\mathbb{I}\{x_n>0\},\ n\in\{2,3,4\};
\end{align*}
ties are broken randomly. 

The above joint-routing-scheduling policy is called GSP, because the PL $Q$ function is a generalized measurement of the ``temporal length'' of each path.
By this policy, server $n$ will prioritize the jobs from class $p$ if server $n$ is the bottleneck of path $p$. If server $n$ is the bottleneck of multiple paths, the GSP policy would further prioritize the jobs from the path with a larger $Q$ function.
We also define
$$P^*=\mathop{\arg\min}\nolimits_{p\in\mathcal{P}}\gamma_p Q_p(x);$$
note that $1\leq |P^*|\leq 3$.

We say that the traffic in the network is $stable$ if there exists $C<\infty$ such that for any initial condition, 
\begin{align*}
    \limsup\limits_{t \rightarrow \infty}\frac{1}{t}\int_{s=0}^t\mathsf E[\|x(s)\|]ds < C.
\end{align*}
We say the network is $stabilizable$ if $\Bar{\lambda}<\Bar{\mu}$, where $\Bar{\mu}$ is the min-cut capacity of the network. Note that $\Bar{\lambda}<\Bar{\mu}$ ensures the existence of a stabilizing Bernoulli routing policy.
\section{Stability guarantee for GSP policy}
\label{sec_stability}

In this section, we consider the stability condition under GSP policy for the bridge network.

To state the main result, let $\mathcal{M}$ be the set of cuts of the network, and define 
\begin{align*}
    m = \quad&\min \quad \frac{\sum_{n\in (M-\tilde{M})}\bar\mu_n}{\bar\lambda-\sum_{n\in\tilde{M}}\bar\mu_n},\\
    & \begin{array}{r@{\quad}l@{}l@{\quad}l}
    s.t.&\bar\lambda-\sum_{n\in\tilde{M}}\bar\mu_n>0,\  \tilde{M}\subsetneq M,\\
    &M\in \mathcal{M};
    \end{array}
\end{align*}
one can show that $m>1$ if the network is stabilizable. Let $d_n$ be the number of links on the longest path from the origin to server $n$. For each $M\in\mathcal M$, let $\mu_{\max}^M=\max\limits_{n\in M}\bar\mu_n$, $\mu_{\min}^M=\min\limits_{n\in M}\bar\mu_n$, and define
\begin{align*}
    &G_1^M := \sum_{n\in\mathop{\arg\max}\limits_{\nu\in M}\bar\mu_\nu}d_n,\quad
    G_2^M := \min_{n\in\mathop{\arg\min}\limits_{\nu\in M}\bar\mu_\nu} d_n,\nonumber\\
    &G^M := \sgn(\mu_{\max}^M-\mu_{\min}^M)(G_2^M-G_1^M)_+,
\end{align*}
where $\sgn(\cdot)$ (resp. $(\cdot)_+$) is the sign (resp. positive part) function.
Also define 
$$
\Delta_G:=\sgn\Big(\max_{M\in\mathcal M}G^M\Big).
$$
Then, we can state the main result as follows:
\begin{theorem}\label{theorem1}
The bridge network is stable under the GSP policy if the network is stabilizable and 
\begin{align}
\label{equa:theorem}
    1<\gamma^{(2+\Delta_G)}<\beta^{(2+\Delta_G)}<m.
\end{align}
\end{theorem}

The above result includes two key points.
First, the GSP policy is stabilizing if the queuing network is stabilizable; i.e., the GSP policy maximizes the throughput of the network. 
Second, constraints on the parameters for the GSP policy are given, which determine the feasible set for the subsequent learning-based optimization (Section~\ref{sec_learning}).

The parameters $m$ and $\Delta_G$ indicate the relationships between the service rates and arrival rate, while the design parameters $\beta$ and $\gamma$ characterize the GSP policy's ``preference''. 
Note that $\lim_{\bar\lambda\uparrow{\bar\mu}} m = 1$, which means each server and each path will be ``equally emphasized'' by the GSP policy as the network approaches saturation. The case when $\Delta_G=1$ indicates that the path with a stronger service ability has a bottleneck closer to the origin. Then the degree of ``preference'' should decrease and the constraints of $\beta$ and $\gamma$ should be stricter because even though with a weaker service ability, the other paths have the bottleneck closer to the destination, and jobs on these paths can get to the destination more quickly once they pass the bottleneck.

The proof of Theorem~\ref{theorem1} uses a Lyapunov function constructed from the PL $Q$ functions:
\begin{equation}
\label{equa:L}
V(x)=\sum_{p\in\mathcal P}\Big(Q_p(x)\Big)^2.
\end{equation}
The key technique is to utilize the connection between the locations of bottlenecks and the GSP policy to argue for the negative mean drift of the Lyapunov function. In particular, we show that there exist $\epsilon > 0,~C < \infty$ such that for all $x \in \mathcal{X}$
\begin{equation}
\label{equa:drift}
\mathcal{L}V({x})=-\epsilon \|x\|_1 + C,
\end{equation}
where $\mathcal L$ is the infinitesimal generator of the network model under the GSP policy.
\eqref{equa:drift} implies the boundedness of the 1-norm of the traffic state and thus the stability of the network, according to the Foster-Lyapunov criterion \cite{meyn_tweedie_1993}.

\emph{Proof of Theorem \ref{theorem1}.}
Firstly, we consider the case when $Q_p = 0$ for some $p\in\mathcal{P}$. Suppose $\mathcal{P}^*=\{(1,2)\}$ and $Q_{12} = 0$, which implies the servers on path $(1,2)$ are all empty. According to the GSP policy, $\lambda_{12}(x)=\bar{\lambda},~\lambda_{135}(x)=\lambda_{45}(x)=0$. The jobs will be allocated to path $(1,2)$, thus it only makes finite positive contribution to the drift. While path $(1,3,5)$ and $(4,5)$ are not empty and keep serving jobs, they make negative contribution to the drift:
  \begin{equation}
  \begin{aligned}
  \mathcal{L}V(x)&=-\epsilon_{135}\sum_{n \in (1,3,5)}x_n^{135}-\epsilon_{45}\sum_{n \in (4,5)}x_n^{45}+C \\&< -\epsilon\sum_{n \in \mathcal{N}}x_n+C.
  \end{aligned}
  \end{equation}
Thus we can infer that the Lyapunov drift of the network can only be negative or finite positive when there are empty paths in the network, which also covers the cases of $Q_{135}=0$ and $Q_{45}=0$.

The rest of this proof is devoted to the case where $Q_p>0, p\in\mathcal{P}$. Specifically, we consider five qualitatively different cases:
\\\underline{Case 1:} $B=(2,5,5)$.
\par In this case, the $Q$ function of each path have the forms of:
\begin{align*}
    &Q_{12} =\beta(x_1^{12}+x_2^{12}),\nonumber\\
    &Q_{135} = (x_1^{135}+x_3^{135}+x_5^{135}),\nonumber\\
    &Q_{45} =\beta(x_4^{45}+x_5^{45}).
\end{align*}
Then we prove according to the different routing decision:
\begin{enumerate}
    \item [\romannumeral 1)] $P^* = \{(1,2)\}$
    \par Since $b_{135} = b_{45} = 5,~b_{12} = 2$, we have:
    \begin{align*}
        \gamma_{12}(B) = 1,\gamma_{135}(B) = \gamma_{45}(B) = \gamma&,\bar\mu_{b_{12}}>\bar\mu_{b_{135}}=\bar\mu_{b_{45}}\nonumber\\
        \gamma_{12}(B) = \gamma,\gamma_{135}(B) = \gamma_{45}(B) = 1&,\bar\mu_{b_{135}}=\bar\mu_{b_{45}}>\bar\mu_{b_{12}}.
    \end{align*}
    Then the new arriving job will go to path $(1,2)$ when: 
    \begin{subequations}
    \begin{align}
        &1)~\gamma Q_{12}<Q_{135},Q_{45};\label{equa:B225-r12_a}\\
        &2)~Q_{12}<\gamma Q_{135},\gamma Q_{45}.\label{equa:B225-r12_b}
    \end{align}
    \end{subequations}
     The scheduling decision of server 1 and server 5 will be $\mu_2^{12}(x) = \bar\mu_2$ and $\mu_5^{45}(x)+\mu_5^{135}(x) = \bar\mu_5,~\mu_5^{45}(x)\mu_5^{135}(x)=0$. Clearly (\ref{equa:B225-r12_a}) can be transformed into $Q_{12}<Q_{135},Q_{45}$, then we have :
    \begin{align*}
        \mathcal{L}V(x)<&2\beta^2(\lambda-\mu_2^{12}(x))(x_1^{12}+x_2^{12})\nonumber\\
        &-2\beta^2\mu_5^{45}(x)(x_4^{45}+x_5^{45})\nonumber\\
        &-2\mu_5^{135}(x)(x_1^{135}+x_3^{135}+x_5^{135}) + C\nonumber\\
        <&2\beta(x_1^{12}+x_2^{12})\Big(\beta(\bar\lambda-\bar\mu_2)-\bar\mu_5/\gamma\Big)+C.
    \end{align*}
    The drift is naturally negative when $\bar\lambda-\bar\mu_2\leq0$, and it remains negative since $1<\beta\gamma<\beta^2<m \leq \frac{\bar\mu_5}{\bar\lambda-\bar\mu_2}$ when $\bar\lambda-\bar\mu_2>0$ according to (\ref{equa:theorem}). Thus (\ref{equa:drift}) holds.
    \item [\romannumeral 2)] $P^* = \{(1,3,5)\}$
    \par Similarly, the jobs will go to path (1,3,5) when $\gamma Q_{135}<Q_{12},\gamma Q_{45}$ or $Q_{135}<\gamma Q_{12},Q_{45}$. We have the drift:
    \begin{align*}
        \mathcal{L}V(x)<&2(x_1^{135}+x_3^{135}+x_5^{135})(\bar\lambda-\beta\bar\mu_2/\gamma-\beta\bar\mu_5)\nonumber\\
        &+C.
    \end{align*}    
    \item [\romannumeral 3)] $P^* = \{(4,5)\}$
    \begin{align*}
        \mathcal{L}V(x)<& 2\beta(x_4^{45}+x_5^{45})(\beta\lambda-\beta\bar\mu_2/\gamma-\bar\mu_5)+C.
    \end{align*}    
    Since $\gamma<\beta<\sqrt{m}$, the drift satisfies (\ref{equa:drift}).
\end{enumerate}
For other cases when there are ties and $|P^*|>1$, since a job can only take a single path, the cases can be proved analogously.
\\\underline{Case 2:} $B=(1,5,5)$.
\par When $P^* = \{(1,2)\}$, the maximal coefficient of $x_1^{12}$ exists when $\bar\mu_1>\bar\mu_5$ and $Q_{12}<\gamma Q_{135},\gamma Q_{45}$, the drift turns to be:
    \begin{align*}
        \mathcal{L}V(x)<&2\beta^2x_1^{12}\Big(\beta^2(\bar\lambda-\bar\mu_1)-\bar\mu_5/\gamma\Big)+C.
    \end{align*}
     Since $d_1=1, d_5=3$ and the two nodes belong to the same cut, we have $\Delta_G = 1$ and $1<\gamma<\beta<m^{\frac{1}{3}}$, which make (\ref{equa:drift}) stands.
    The proof of other forms of $P^*$ is analogous with case 1.
\\\underline{Case 3:} $B=(2,5,4)$.
\par We only show the case where $P^* = \{(1,2)\}$, the other cases can be proved analogously. For the diverse possible values of $\gamma_p(B)$, the drift has the form of:
\begin{align*}
    \mathcal{L}V(x)<&2\beta(x_1^{12}+x_2^{12})\Big(\beta(\bar\lambda-\bar\mu_2)-\bar\mu_5\frac{\gamma_{12}(B)}{\gamma_{135}(B)} \nonumber\\
    &-\beta^2\bar\mu_4\frac{\gamma_{12}(B)}{\gamma_{45}(B)}\Big)+C.
\end{align*}
According to the GSP policy, we have the scaled ratio:
\begin{align*}
    &\Big(\frac{\gamma_{12}(B)}{\gamma_{135}(B)},\frac{\gamma_{12}(B)}{\gamma_{45}(B)}\Big) \in \nonumber\\
    &\Big\{\Big(1,1\Big),\Big(\frac{1}{\gamma^2},\frac{1}{\gamma}\Big),\Big(\frac{1}{\gamma},\frac{1}{\gamma^2}\Big),\Big(\frac{1}{\gamma},1\Big),\Big(1,\frac{1}{\gamma}\Big),\Big(\frac{1}{\gamma},\frac{1}{\gamma}\Big)\Big\}.
\end{align*}
Thus the possible maximal coefficient of $x_1^{12}+x_2^{12}$ exists when:
\begin{align*}
    1)&\bar\mu_2>\bar\mu_4>\bar\mu_5,~\Big(\frac{\gamma_{12}(B)}{\gamma_{135}(B)},\frac{\gamma_{12}(B)}{\gamma_{45}(B)}\Big)=\Big(\frac{1}{\gamma^2},\frac{1}{\gamma}\Big),\nonumber\\
    &Q_{12}<\gamma^2Q_{135},\gamma Q_{45};\nonumber\\
    2)&\bar\mu_2>\bar\mu_5>\bar\mu_4,~\Big(\frac{\gamma_{12}(B)}{\gamma_{135}(B)},\frac{\gamma_{12}(B)}{\gamma_{45}(B)}\Big)=\Big(\frac{1}{\gamma},\frac{1}{\gamma^2}\Big),\nonumber\\
    &Q_{12}<\gamma Q_{135},\gamma^2 Q_{45}.
\end{align*}
We have:
\begin{align*}
    \mathcal{L}V(x)<&C+2\beta(x_1^{12}+x_2^{12})\Big(\beta(\bar\lambda-\bar\mu_2)\nonumber\\
    &+\mathbb{I}_{\{\bar\mu_2>\bar\mu_4>\bar\mu_5\}}(-\bar\mu_5/\gamma^2-\beta^2\bar\mu_4/\gamma)\nonumber\\
    &+\mathbb{I}_{\{\bar\mu_2>\bar\mu_5>\bar\mu_4\}}(-\bar\mu_5/\gamma -\beta^2\bar\mu_4/\gamma^2)\Big).
\end{align*} 
Since there are $d_2 = 2,d_4 = 1, d_5 = 3$ and $\Delta_G = 1$ when $\bar\mu_2>\bar\mu_4>\bar\mu_5$, we have $1<\beta\gamma^2<m$. Thus the (\ref{equa:drift}) can be proved. The case when $B=(1,5,4)$ can be proved analogously. 
\\\underline{Case 4:} $B=(2,3,4).$
\par We discuss when $P^* = \{(4,5)\}$. When the possible maximal coefficient of $x^{45}_4$ exists, the drift is:
\begin{align*}
    \mathcal{L}V(x)<&C+\beta x^{45}_4 \Big(\beta(\bar\lambda-\bar\mu_4)\nonumber\\
    &+\mathbb{I}_{\{\bar\mu_4>\bar\mu_3>\bar\mu_2\}}(-\bar\mu_2/\gamma^2-\bar\mu_3/\gamma)\nonumber\\
    &+\mathbb{I}_{\{\bar\mu_4>\bar\mu_2>\bar\mu_3\}}(-\bar\mu_2/\gamma-\bar\mu_3/\gamma^2)\Big).
\end{align*}
Since $d_2 = 2,d_4 = 1, d_3 = 2$ and $\Delta_G = 1$ when $\bar\mu_4>\bar\mu_2>\bar\mu_3$ or $\bar\mu_4>\bar\mu_3>\bar\mu_2$, thus we have $1<\beta\gamma^2<\beta^3<m$ and (\ref{equa:drift}) holds. The cases of other forms of  $P^*$ when $B=(2,3,4)$ and $B=(2,3,5)$ can be proved analogously.
\\\underline{Case 5:} $B=(1,3,5).$
\par Considering $P^* = \{(1,2)\}$, when the possible maximal coefficient of $x^{12}_1$ exists, the drift satisfies:
\begin{align*}
    \mathcal{L}V(x)<&C+\beta^3 x^{12}_1\Big(\beta(\bar\lambda-\bar\mu_1)\nonumber\\
    &+\mathbb{I}_{\{\bar\mu_1>\bar\mu_5>\bar\mu_3\}}(-\bar\mu_3/\gamma^2-\bar\mu_5/\gamma)\nonumber\\
    &+\mathbb{I}_{\{\bar\mu_1>\bar\mu_3>\bar\mu_5\}}(-\bar\mu_3/\gamma-\bar\mu_5/\gamma^2)\Big).
\end{align*}
We have $d_1 = 1,d_3= 2,d_5 = 3$, and there is $\Delta_G = 1$ when $\bar\mu_1>\bar\mu_3>\bar\mu_5$. Thus there is $1<\beta\gamma^2<\beta^3<m$, and (\ref{equa:drift}) stands. The other forms of $P^*$ of case $B=(1,3,5)$ and $B=(1,3,4)$ can be proved analogously.
The proof is the same for other cases of $B=(1,1,4)$, $B=(1,1,5)$, $B=(2,1,4)$, $B=(2,1,5)$.
\par In conclusion, we show that the mean drift always satisfies (\ref{equa:drift}), which ensures the stability of the network.
\hfill$\square$

\section{Iterative computation of GSP parameters}
\label{sec_learning}

In this section, we develop a policy iteration (PI) algorithm that computes the parameters for the GSP policy. In particular, theorem~\ref{theorem1} gives a set of values for $\beta$ and $\gamma$ that ensure stability; the goal of the PI algorithm is to search the aforementioned set for a specific combination of $\beta$ and $\gamma$ that attains near-optimal traffic efficiency. In addition, the PI algorithm does not require a priori knowledge of the arrival and service rate. We also compare the performance of the GSP PI algorithm with three benchmark policies, viz. neural network (NN)-based PI, simple shortest path, and optimized Bernoulli. 

We formulate the joint routing-scheduling problem as a Markov decision process (MDP) with state space $\mathcal X=\mathbb Z_{\ge0}^7$. 
Since the routing and scheduling decision will only be made when an arrival or a departure occurs (i.e., at transition \emph{epochs} \cite[p.72]{gallager2013stochastic}), the problem can be formulated as a discrete-time (DT) MDP.
We denote the state and action of the DT MDP as $x[k]$ and $a[k]$, respectively.
With a slight abuse of notation, $x[k]=x(t_k)$, where $t_k$ is the $k$-th transition epoch of the continuous-time process.
Similarly, the action is given by $a[k]=(\lambda[k],\mu[k])$, where $\lambda[k]\in\{0,\bar\lambda\}^3$ and $\mu[k]\in\prod_{n\in\mathcal N}\{0,\bar\mu_n\}^{m_n}$ are the route-specific arrival rates and service rates, respectively.

As indicated in Section~\ref{sec_model}, both the routing and the scheduling decisions can be parameterized via design variables $\beta$ and $\gamma$; by Theorem~\ref{theorem1}, $\beta$ and $\gamma$ should satisfy \eqref{equa:theorem}.
Thus, the action is given by the policy
$$
a[k]=\pi(x[k];\beta,\gamma).
$$
The transition probability $p(x'|x,a)$ of the DT process can be derived from the network model defined in Section~\ref{sec_model} in a straightforward manner.
The one-step cost of the MDP is given by
$$
u[k]=\|x[k]\|_1(t_k-t_{k-1}).
$$
The total cost over one episode of the MDP is thus given by
$$
U_\pi^K(x)=\mathrm{E}_\pi\Big[\sum_{\ell=k}^K\|x[\ell]\|_1(t_k-t_{k-1})\Big|x[0]=x\Big],
$$
where $K\in\mathbb Z_{>0}$ is the length (i.e., number of transition epochs) of one episode.

\subsection{GSP PI algorithm}
Closed-form solution to $U_\pi^K$ is not easy. Hence, we use the PL $Q$ function defined by \eqref{equa:L} as a proxy for $U_\pi^K$.
A key characteristic of our formulation is that $(\beta,\gamma)$ parametrizes both the policy $\pi(x;\beta,\gamma)$ and the PL functions $\gamma_pQ_p(x;\beta)$. Although this may lead to a gap towards the ground-truth optimal solution, it significantly simplifies the computation and, importantly,
guarantees traffic stability.

The GSP PI algorithm has two tasks in parallel: calibration of model data $\bar\lambda$, $\bar\mu_n$ and optimization of control parameters $\beta,\gamma$. The latter depends on the former, since $\Delta_G$ in \eqref{equa:theorem} depends on model data. Let $\hat\lambda^{(i)},\hat\mu_n^{(i)}$ be the estimate of $\bar\lambda,\bar\mu_n$ after $i$ iterations (i.e., episodes), $\Delta_G^{(i)}$ be the corresponding $\Delta_G$, and $\beta^{(i)},\gamma^{(i)}$ be the control parameters after $i$ iterations.
Then, the specific steps of the GSP PI algorithm
can be summarized as follows:
\begin{enumerate}
    \item {\bf Initialize} with $\hat\lambda^{(0)}\in\mathbb R_{>0}$, $\hat\mu_n^{(0)}\in\mathbb R_{>0}$ for each server $n$, $\beta^{(0)}\in(1,m^{\frac{1}{(2+\Delta_G)}})$, $\gamma^{(0)}\in(1,\beta^{(0)})$.
    
    \item {\bf Policy evaluation}: Simulate a Monte-Carlo episode of the network model with the policy $\pi(x;\beta^{(i-1)},\gamma^{(i-1)})$. Update the estimates to obtain $\hat\lambda^{(i)},\hat\mu_n^{(i)}.$ Update the parameters $\beta^{(i)},\gamma^{(i)}.$ (See below for how the updates are done.)
    
    \item {\bf Policy improvement}: Use $\beta^{(i)},\gamma^{(i)}$ to obtain a new policy $\pi(x;\beta^{(i)},\gamma^{(i)})$.
    \item {\bf Terminate} when $|\beta^{(i)}-\beta^{(i-1)}|$ and $|\gamma^{(i)}-\gamma^{(i-1)}|$ are smaller than some thresholds.
\end{enumerate}

The rest of this subsection is devoted to the update of $(\beta,\gamma).$
Given $(\beta,\gamma)$, suppose that a job arrives at epoch $k$ and takes path $p$ according to policy $\pi(x;\beta,\gamma)$. Then, we track the system time $W[k]$\footnote{This notation should cause no confusion, since at most one Poisson arrival can occur at an epoch.} of this particular job, and use the square error
$$
\mathrm{SE}_{(\beta,\gamma)}[k]=\mathbb I_{\{W[k]>0\}}\Big(\gamma_{p[k]}Q_{p[k]}(x[k];\beta)-W[k]\Big)^2
$$
to update $(\beta,\gamma)$, where $p[k]$ is the path that $\pi(x[k];\beta,\gamma)$ selects. Note that $W[k]$ records the time that a job spends in the continuous-time setting; we define $W[k]=0$ if no arrival occurs at epoch $k$, so $\mathbb I_{\{W[k]>0\}}$ indicates whether there is an arrival at epoch $k$. $(\beta,\gamma)$ are obtained by minimizing the sum-of-squares error: 
\begin{align}
    (\beta^*,\gamma^*)=\underset{{\footnotesize(\beta,\gamma)\mbox{ satisfy }\eqref{equa:theorem}}}{\mbox{arg min}}\sum_{k=1}^K\mathrm{SE}_{(\beta,\gamma)}[k].
    \label{eq_beta*}
\end{align}
Note that the PL function actually approximates the job-specific system time $W$ rather than the total system time $U_\pi^K$, which are in fact related via
$$
\lim_{K\to\infty}\Bigg\{\frac{U_\pi^K(x)}{t_K}-\frac{\sum_{k=1}^KW[k]}{\sum_{k=1}^K\mathbb I_{\{W[k]>0\}}}\Bigg\}=0
\quad a.s..
$$

\subsubsection{Estimate of $\bar\lambda,\bar\mu_n$}
In the $i$th iteration, we run a Monte-Carlo simulation of the network model. The simulation has a discrete time step size that is sufficiently small to approximate the continuous time process; i.e., we use Bernoulli processes with sufficiently small time increment to approximate Poisson processes.
We track every inter-arrival time at the origin and compute the average to obtain $1/\hat\lambda^{(i)}.$ We also track every service time at each server $n$ and compute the average to obtain $1/\hat\mu_n^{(i)}.$

\subsubsection{Update of $\beta$}
In the $i$th iteration, we record $\{x[k],p[k],W[k]\}$ if $W[k]>0$. 
Let $D^{(i)}$ be the table of $\{x[k],p[k],W[k]\}$.
We determine $\beta^{(i)}$ by a method similar to the {least-squares partition algorithm} in \cite{Magnani2009}. We first partition $D^{(i)}$ according to the boundaries of the PL functions $\{Q_p(x;\beta^')\}$ with $\beta^'=\beta^{(i-1)}$.
Then we use the L-BFGS-B algorithm \cite{zhu1997algorithm} to calculate a better $\beta'$ that minimizes the sum-of-squares error defined in \eqref{eq_beta*}, with $\gamma$ fixed to be $\gamma^{(i-1)}$. This leads to a new partition according to $\beta'$. We recursively do so until the partition stabilizes or the iteration reaches a certain maximum number, and the last $\beta'$ is taken as $\beta^*$ of this episode.

\subsubsection{Update of $\gamma$}
After $\beta^{(i)}$ is obtained, we may determine the relation between $\gamma_p$ and $\gamma$ for each set of data in $D^{(i)}$. To be specific, we determine the combination of bottlenecks $B$ at state $x[k]$ based on $\beta^{(i)}$, and $\gamma_p(B)$ is given by \eqref{path gamma}. Then, we again use the L-BFGS-B algorithm to calculate $\gamma^*$ which minimize the sum-of-squares error defined in \eqref{eq_beta*} with $\beta$ fixed to be $\beta^{(i)}$. When update of $\gamma$ is completed, $\gamma^*$ of this iteration is then set to be $\gamma^{(i)}$. The process of the $i$th iteration of $\gamma$ and $\beta$ is given by Algorithm \ref{beta gamma calibration}.

\begin{algorithm} 
    \caption{Computation of $\beta^{(i)}$ and $\gamma^{(i)}$}    
    \label{beta gamma calibration}       
    \begin{algorithmic}[1] 
    \Require $D^{(i)},\hat\lambda^{(i)},\hat\mu_n^{(i)},\beta^{(i-1)},\gamma^{(i-1)}$     
    \State partition data with $\beta^'\longleftarrow\beta^{(i-1)}$   
    \Repeat 
    \State update $\beta^' $ based on current partition   \Until{partition stabilizes}
    \State $\beta^{(i)} \longleftarrow \beta^*$
    \State determine the forms of $\gamma_p$ for data in $D^{(i)}$
    \State $\gamma^{(i)} \longleftarrow \gamma^*$
    \Ensure  $\beta^{(i)}$ , $\gamma^{(i)}$
    \end{algorithmic} 
\end{algorithm}

\subsection{Benchmark policies and algorithms}
\par To evaluate the performance of the GSP PI algorithm, we consider three benchmarks:
\subsubsection{Neural network (NN) PI}
We consider action-value-function $\Theta(x,a)$ and use NN to approximate it. Initially, we get the first sequence of state-action-reward $\{x[k],a[k],r[k]\}$ and 
\begin{align*}
    \theta_0(x[k],a[k])=\sum_{k'=k}^K l^{(k'-k)}r_{k'},~\Theta_0=\theta_0
\end{align*}
for the initial iteration after a Monte-Carlo episode, where $l$ is a discount of future action-state-reward. The value is renewed as 
\begin{align*}
    \Theta_i(x[k],a[k])=&\Theta_{i-1}(x[k],a[k])+(\theta_i(x[k],a[k])\\
    &-\Theta_{i-1}(x[k],a[k]))/i
\end{align*}
in iteration $i,i\in N^*$. Then the current policy is evaluated by a designed NN, which has two convolutional layers, two maxpooling layers and two fully connected layers. We adopt the rectified linear unit (ReLU) activation function at each layer to avoid gradient explosion and extinction. The weights of NN are updated by the adaptive moment estimation (Adam) algorithm \cite{Kingma2014AdamAM} to narrow the gap between the predicted and calculated state-action-value. Finally, we use the NN approximation for $\Theta$ for policy improvement by the $\epsilon$-greedy algorithm. Since an exact optimal policy of the original MDP is not readily available, the policy computed by NN PI is used as an approximate optimal policy.

\subsubsection{Simple shortest-path (SSP) policy}
For routing and scheduling decisions under SSP policy, we simply calculate the cost of paths by $\Phi(x,p)=\sum_{n\in p}x^p_n$ and select the path with the smallest $\Phi(x,p)$ value.

\subsubsection{Optimal Bernoulli (OB) policy}
A Bernoulli policy route any incoming job to a route $p\in\mathcal P$ with a time-invariant probability $\eta_p$ such that
$
\sum_{p\in\mathcal P}\eta_p=1.
$
Consequently, the routing probabilities $\eta_p$ are optimized using a simple numerical search scheme.

\subsection{Evaluation and comparison}

We compare our GSP policy with the three benchmarks according to the following performance metrics:
\begin{enumerate}
    \item [1)] {\bf Computational efficiency}: training time and number of iterations (if applicable.)
    \item [2)] {\bf Implementation efficiency}: the average system time experienced by all jobs that go through the network.
\end{enumerate}

Consider the network in Fig 1 and suppose $\bar\mu_1=0.15$, $\bar\mu_2=0.1$, $\bar\mu_3=0.25$, $\bar\mu_4=0.15$, $\bar\mu_5=0.2$, $\bar\lambda=0.2$, all in unit sec$^{-1}$. Then we have $m=1.50,~\Delta_G=0 $ for this set of parameters.
For GSP PI, we initialize the estimates of $\bar\mu_n$ with 0.5 and estimate of $\bar\lambda$ with 0.1.
The OB policy turns out to be $\eta^*=[0.28, 0.20, 0.52]^T$.
For simulation, a discrete time step of 0.1 sec is used. All experiments were implemented in Google Colab \cite{colab}, using Intel(R) Xeon(R) CPU with 12.7GB memory. NN PI was trained on NVIDIA GeForce RTX 2080 Ti with 40GB memory.

\begin{figure}[thpb]
\centering
\label{fig:comp_learn}
\subfigure[The time of learning]{
\begin{minipage}[t]{1\linewidth}
\centering
\includegraphics[width=1\textwidth]{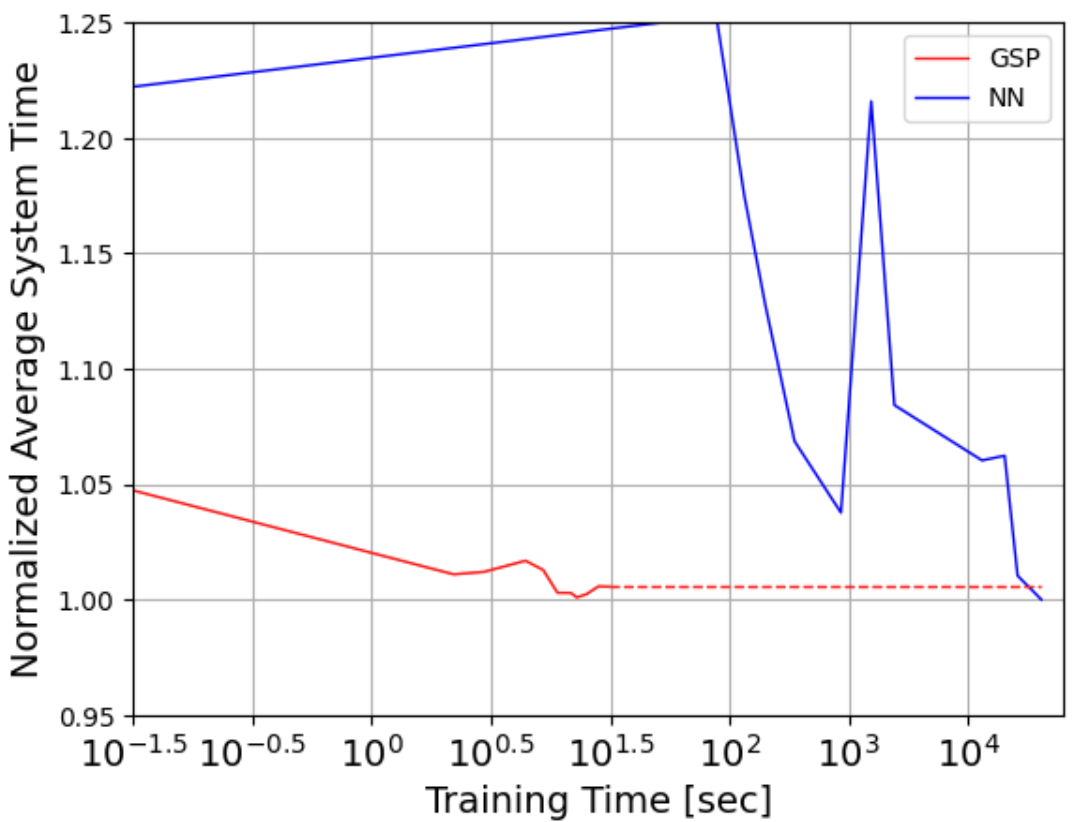}
\label{fig:comp_learnTime}
\end{minipage}%
}%
\quad
\subfigure[The number of iterations]{
\begin{minipage}[t]{1\linewidth}
\centering
\includegraphics[width=1\textwidth]{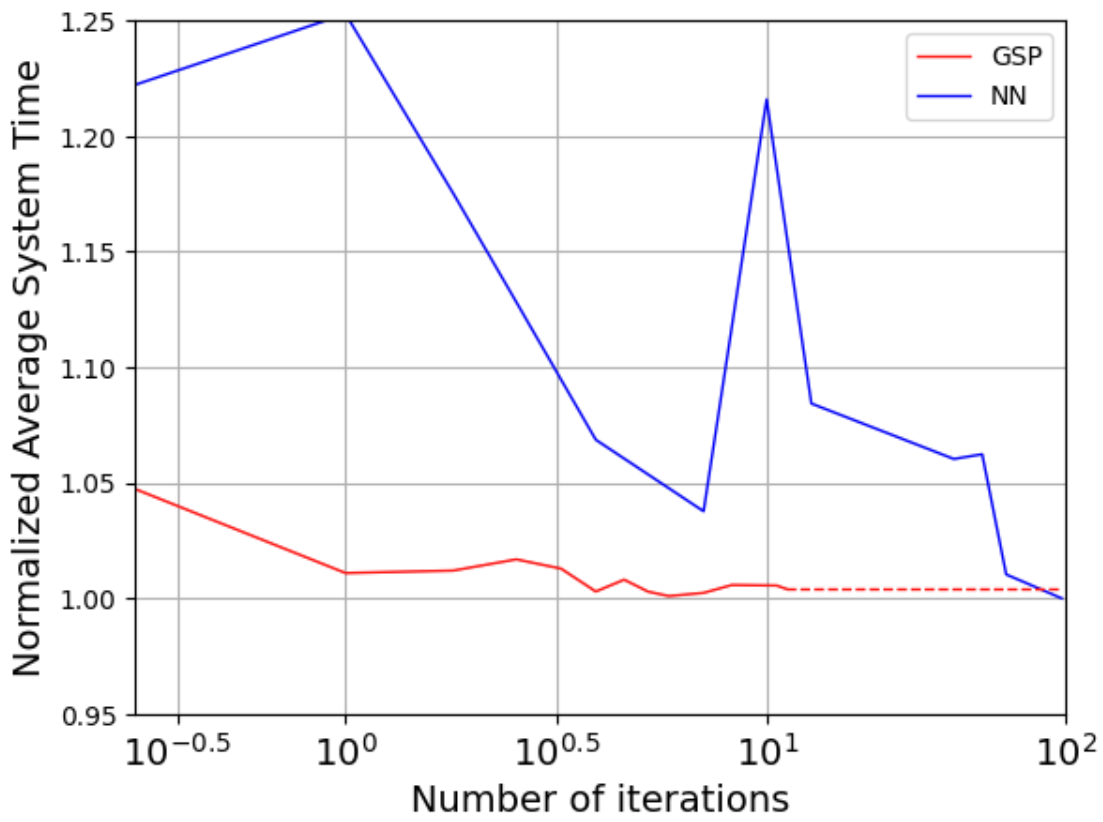}
\label{fig:comp_learnIteration}
\end{minipage}%
}%
\centering
\captionsetup{font={small}}
\caption{Average system time comparison of GSP PI and NN algorithm. The coordinate scale of the axes is uneven due to the large value. (a) Average system time as the learning time growing. (b) Average system time as the number of iterations increases.}
\end{figure}

The computational efficiency of our GSP policy is significantly better than the general NN algorithm with a 97.8\%-99.8\% reduction of training time as demonstrated in Fig. \ref{fig:comp_learnTime}. 
Our GSP PI algorithm converges rapidly within a few seconds while NN PI takes more than twelve hours to converge. This is expected, since the PL functions have simple structures, while the NN model require sophisticated training.

Table \ref{table:ave_job_T_4methods} lists the Normalized Average System Times (NAST) under various methods. The results are generated from test simulations of $5\times10^6$ sec.
Although NN PI performs better in long terms of learning as expected, GSP PI performs better with just a few number of iterations as demonstrated in Fig. \ref{fig:comp_learnIteration}, and can attain a 1.9\%-6.7\% optimality gap towards NN PI. 
\begin{table}[!htb]
    \centering
    \caption{Normalized Average System Times (NAST) of various schemes}
    \label{table:ave_job_T_4methods}
    \begin{tabular}{l|l}
        \hhline
        Algorithm & NAST\\ \hhline
        Neural network (NN)  & 1.00\\ \hhline
        Generalized shortest-path (GSP) & 1.03\\ \hhline
        Simple shortest-path (SSP) & 1.12\\ \hhline
        Optimal Bernoulli (OB) & 1.25\\ 
        \hhline
    \end{tabular}
\end{table}
The job will spend more time going through the queuing network under SSP or OB policy at the average of $35.04$ sec and $39.27$ sec. Though the implementation efficiency of GSP PI is slightly worse than NN PI, GSP PI gives the best trade-off between computational efficiency and implementation efficiency: the average system time of NN and GSP method are $31.30$ sec and $32.24$ sec, which implies an adequate optimality gap of 3\%.
\section{Concluding remarks}
\label{sec_conclude}
In this article, we focus on a class of control policies for a single-origin-single-destination network, which we call the generalized shortest-path policy. We design a set of piecewise-linear functions as a proxy of the travel cost on the path over the network. We use the piecewise-linear functions to construct a Lyapunov function to derive theoretical guarantee on the stability of the traffic state. We also develop a policy iteration algorithm to learn the parameters for the proposed policy. Then we implement and compare the algorithm with three other benchmarks including a standard neural network-based algorithm. The experimental results showed that our policy and PI algorithm can perform efficiently and steadily when significantly reducing the computation cost as well, which is cost-effective.
Future work may include (i) extension of the theoretical results as well as PI algorithm to general single-origin-single-destination acyclic networks and (ii) refinement of the PI algorithm to develop more efficient temporal-difference algorithms.
\bibliographystyle{unsrt}
\bibliography{references}
\balance
\end{document}